\documentclass[preprint]{revtex4}
\usepackage{mathrsfs}
\usepackage{epsfig,bm,dcolumn}
\usepackage{graphicx}
\usepackage{color}
\usepackage{amsmath}

\begin{document}


\title{Molecular Bose-Einstein condensation in a Bose gas with a wide Feshbach resonance at finite temperatures}

\author{Zeng-Qiang Yu and Lan Yin}
\email{yinlan@pku.edu.cn}
\address{School of Physics, Peking University, Beijing 100871, China}

\date{\today}

\begin{abstract}

Bose-Einstein condensation (BEC) of Feshbach molecules in a
homogeneous Bose gas is studied at finite temperatures in a single-channel mean-field approach
where the Hartree-Fock energy and pairing gap are determined self-consistently.
In the molecular-BEC state, the atomic excitation is
gapped and the molecular excitation is gapless.  The binding energy
of Feshbach molecules is shifted from the vacuum value due to
many-body effect.  When the scattering length $a_s$ of atoms is
negative, the system is subject to mechanical collapse due to
negative compressibility.  The system is stable in most regions with
positive scattering lengths.  However at low temperatures near the
resonance, the molecular-BEC state vanishes, and the coherent
mixture of atomic and molecular BEC is subject to mechanical
collapse.

\end{abstract}

\pacs{}


\maketitle

\section{Introduction}

Many interesting phenomena have been observed in Bose gases with
Feshbach resonances.  Enormous particle-loss rate due to three-body
recombination was found near the resonance \cite{Ketterle99}.
Oscillations between atoms and diatomic Feshbach molecules were
generated by a sudden change in magnetic field \cite{Wieman02}.
Large number of Feshbach molecules were produced from Bose gases
either by tuning the magnetic field through the resonance
\cite{Ketterle03} or by oscillating the magnetic field at a
frequency corresponding to the molecular binding energy
\cite{Wieman05}.  Even tetramer molecules were created from Feshbach
molecules \cite{Grimm09}.  However BEC of Feshbach molecules in Bose
gases has not been achieved in experiments so far and properties of
this state are waiting to be explored.

Properties of atomic-BEC state have been extensively studied
theoretically \cite{ReviewBEC}.  Phase transition between atomic and
molecular BEC was proposed near the resonance \cite{Radzihovsky04,
Stoof04, Lee04, Braaten07}. However these phases suffer mechanical
collapse in regions with negative scattering length $a_s<0$ \cite{
Mueller05}.  A recent work \cite{Yin08} shows that the molecular-BEC
state of a Bose gas with a wide Feshbach resonance at zero
temperature exists only when the atom density $n$ satisfies
$na_s^3<0.0164$ for positive scattering length $a_s>0$.

In this paper, the molecular-BEC state of a homogeneous Bose gas
with a wide Feshbach resonance is studied at finite temperatures. In
the following, we first describe the molecular-BEC state in a
mean-field approach where the Hatree-Fock energy and pairing gap are
determined self-consistently.  Long-wavelength excitations and molecular
binding energy are obtained.  Then mechanically stability of the
molecular-BEC state is examined across the resonance, and the
mean-field phase diagram is obtained.  Coherent mixture of atomic
and molecular BEC is also studied.  Discussion and conclusion are
given in the end.

\section{Mean-field theory of the molecular-BEC state}

\subsection{Mean-field approach}
In alkali-atom gases with Feshbach resonances, scattering states in
the open channel and bound states in the closed channel are coupled
together.  Near a wide resonance, the effective range of the
interaction is very small, and most atoms are in the open channel
\cite{Strinati, Yin}.  A uniform Bose gas with a wide Feshbach
resonance can be effectively described by a single-channel model,
\begin{align} \label{Hamiltonian}
  \mathcal{H}=-\frac{\hbar^2}{2m}\psi^\dag\nabla^2 \psi+{g\over
  2}\psi^\dag\psi^\dag\psi\psi-\mu \psi^\dag\psi,
\end{align}
where $\psi$ is the field operator of Bose atoms in the open channel,
the coupling constant is given by $g=4\pi\hbar^2a_s/m$, and $\mu$ is
the chemical potential of atoms.

In the molecular-BEC state, atoms are paired into molecules and the
off-diagonal long range order (ODLRO) appears, $\Delta\equiv
g\langle\psi\psi\rangle$.  Without losing any generality,
$\Delta\geq0$ is assumed in the following.  In the mean field,
in addition to the pairing energy,  both Hartree and Fock
energies should be taken into account, and the mean-field
Hamiltonian density is given by
\begin{align} \label{H-mf}
\mathcal{H}_{p}=\mathcal{H}_p^0-\frac{\hbar^2}{2m}\psi^\dag\nabla^2
\psi+(2gn-\mu)\psi^\dag\psi+{\Delta\over 2}(\psi^\dag\psi^\dag
+\psi\psi),
\end{align}
where the constant $\mathcal{H}_p^0$ is given by
$\mathcal{H}_p^0=-(gn^2+\Delta^2/2g)$.  In this mean-field
approximation, the pairing gap $\Delta$ and Hartree-Fock energy
are determined self-consistently, but the
fluctuation and three-body effects are ignored.  The mean-field theory
provides a description about single-particle excitations, but it is
incapable to describe collective excitations or few-body properties
such as Effimov effect.  Nonetheless, the mean-field approximation can
serve as a starting point of the theoretical description for the
molecular-BEC state.

The mean-field Hamiltonian can be diagonalized by Bogoliubov
transformation,
\begin{align} \label{H-Bogo}
H_{p}=\mathcal{E}_0+\sum_{\bf k} E_k \alpha_{\bf k}^\dag\alpha_{\bf
k},
\end{align}
where  $\alpha_{\bf k}=u_k\psi_{\bf k}+v_k\psi_{-\bf k}^\dag$ is the
annihilation operator of atomic quasi-particles,
$E_k=\sqrt{(\epsilon_k-\mu')^2-\Delta^2}$ is the quasi-particle
energy, and the transformation coefficients are given by
$u_k^2=v_k^2+1=\left[1+(\epsilon_k-\mu')/E_k\right]/2$ with
$\epsilon_k=\hbar^2 k^2/2m$ and $\mu'=\mu-2gn$. The energy constant
$\mathcal{E}_0$ is given by
\begin{align}
\mathcal{E}_0={1\over 2} \sum_{\bf k}
\left[E_k-\epsilon_k+\mu'+\frac{\Delta^2}{2E_k}\coth ({\beta
E_k\over2})\right]-gn^2 V,  \nonumber
\end{align}
where $1/\beta=k_{\rm B} T$ and $V$ is the volume.

The order parameter $\Delta$ can be determined self-consistently,
\begin{align}
\Delta= -{g\over V}\sum_{\bf k}u_kv_k[1+2f_k], \nonumber
\end{align}
where $f_k$ is the Bose distribution function of quasi-particles,
$f_k=1/[\exp(\beta E_k)-1]$.  This self-consistency equation can be
written explicitly as
\begin{align} \label{gapeq}
-{m\over 4\pi\hbar^2a_s}=\int \frac{\mathrm{d}^3k}{(2\pi)^3}\left[
\frac{1}{2E_k}\coth \left({\beta E_k\over2}\right)-{1\over
2\epsilon_k} \right],
\end{align}
where the term $-1/2\epsilon_k$ in the integrand on the right-hand
side is a counter term due to vacuum renormalization.  The chemical
potential $\mu$ and the order parameter $\Delta$ can be solved from
Eq. (\ref{gapeq}) and the following equation for the density $n$,
\begin{align} \label{deneq}
n&={1\over V}\sum_{\bf k}\left[v_k^2+(u_k^2+v_k^2)f_k\right]
\nonumber \\ &=\int \frac{\mathrm{d}^3k}{(2\pi)^3}\left[
\frac{\epsilon_k-\mu'}{2E_k}\coth \left({\beta
E_k\over2}\right)-{1\over 2}\right].
\end{align}

\subsection{Excitations in the long-wavelength limit}
There is a gap in the atomic excitation energy given by
\begin{equation}
E_0=\sqrt{\mu'^2-\Delta^2}.
\end{equation}
If the molecular-BEC state is stable, one necessary condition is
that the gap $E_0$ is real, or $-\mu'\geq\Delta$.  When
$-\mu'=\Delta$, the atomic excitation becomes gapless, $E_0=0$,
which marks the lower boundary of the molecular-BEC phase.  The
upper phase boundary is located at the superfluid transition
temperature $T_p$ where $\Delta=0$ and $E_0=-\mu'$.

In addition to atomic excitations, there are collective excitations
of Feshbach molecules.  Although the molecular excitation energy
cannot be obtained in the mean-field approximation, it can be
extracted from poles of two-particle correlation functions given by
\begin{equation}
\chi_{\alpha\beta}({\bf r}-{\bf r}',\tau-\tau') \equiv -{1 \over
\hbar} \langle \mathcal{T}[ b_\alpha({\bf r},\tau) b_\beta^\dagger({\bf
r}',\tau')] \rangle,
\end{equation}
where $b_1=\psi^2$, $b_2=b_1^\dagger$, $b_3=2 \psi^\dagger \psi$,
$0\leq\tau\leq\beta\hbar$, and $\mathcal{T}$ is the time-ordering
operator. As in the zero-temperature case \cite{Yin08}, the
correlation function can be calculated in the random-phase
approximation (RPA),
\begin{equation}\label{cdk}
\chi({\bf k},\omega)=[1-g \chi^{(0)}({\bf k},\omega)]^{-1}
\chi^{(0)}({\bf k},\omega),
\end{equation}
where $\chi^{(0)}({\bf k},\omega)$ is the correlation function
calculated in the mean-field approximation. In RPA, the dispersion
of Feshbach molecules satisfies the following equation,
\begin{equation}\label{pole}
\det|{\rm I}-g\chi^{(0)}({\bf k},\omega)|=0,
\end{equation}
where ${\rm I}$ is the identity matrix.  The molecular excitation is
gapless in the long-wavelength limit, ${\bf k}=0$ and $\omega=0$,
following Eq. (\ref{gapeq}).  At small $k$ and $\omega$, to the
leading order of $k$, the molecular excitation frequency is linearly
dispersed.  The detail of molecular excitations will be studied in
our future work.

\subsection{Binding energy}

\begin{figure}
\includegraphics[width=8cm]{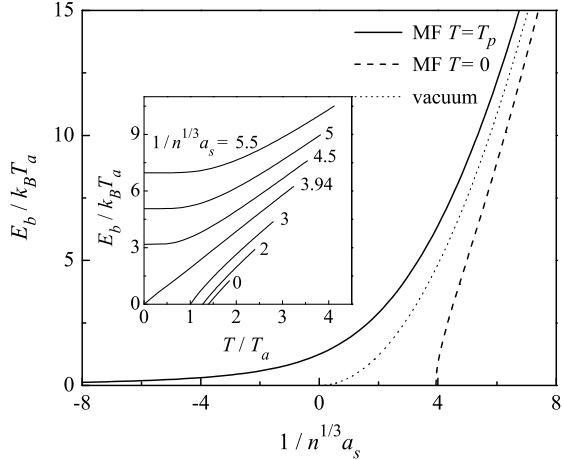}
\caption{Binding energy of Feshbach molecules in molecular-BEC
state. The solid and dashed lines are binding energies at $T=T_p$
and $T=0$K. The dotted line is the binding energy in vacuum,
$E_{b0}=\hbar^2/(ma_s^2)$. The inset shows the temperature
dependence of the binding energy for different
$n^{1/3}a_s$.}\label{bind-energy}
\end{figure}

The binding energy $E_b$ of a Feshbach molecule in the molecular-BEC
state is defined as the energy difference between two atomic and one
molecular excitations at ${\bf k}=0$.  Since the molecular
excitation energy is gapless in the molecular-BEC state, the binding
energy of Feshbach molecules is twice the energy gap of atomic
excitations, $E_b=2E_0$. Due to many-body effect, the binding energy
$E_b$ is different from its vacuum value $E_{b0}=\hbar^2 /(ma_s^2)$.
At the superfluid transition temperature $T_p$, the binding energy
is given by $E_b(T_p)=-2\mu'$.  In the limit of weakly-attractive
interaction, $n^{1/3}a_s\rightarrow 0^-$, we obtain $$E_b(T_p)=16\pi
\zeta^{-{2/ 3}}({3/2})n^{2 \over 3}|a_s|^2k_{\rm B}T_a,$$ where
$\zeta(x)$ is the Riemann-zeta function and $T_a$ is the ideal BEC
temperature, $T_a=(2\pi\hbar^2/mk_{\rm B})[n/\zeta(3/2)]^{2/3}$; in
the limit of weakly repulsion, $n^{1/3}a_s\rightarrow 0^+$, the
binding energy is same as in vacuum,
$E_b(T_p)=E_{b0}=\hbar^2/(ma_s^2)$.

At finite temperatures, molecules not only exist for $a_s>0$, but
also appear as loosely-bound pairs for $a_s<0$ in the mean-field
theory, whereas in vacuum Feshbach molecules exist only for $a_s>0$.
However at zero temperature, the molecular-BEC state exist only when
$1/(n^{1/3}a_s)>3.94$ \cite{Yin08}. In Fig. \ref{bind-energy}, the
molecular binding energy in the mean-field approximation is plotted
at both $T=T_p$ and $T=0$, in comparison with its vacuum value. The
temperature dependence of the binding energy for different
scattering lengths is shown in the inset. For $1/(n^{1/3}a_s)<3.94$,
the binding energy decreases monotonically with temperature, all the
way to zero at the lower boundary of the molecular-BEC phase; for
$1/(n^{1/3}a_s)>3.94$, the binding energy decreases monotonically to
a finite value at zero temperature. In the dilute limit,
$n^{1/3}a_s\rightarrow0^+$, the temperature dependence of the
binding energy is weak, and the binding energy is approximately
given by vacuum value $E_{b0}$.

\subsection{Mechanical stability}

A mechanically stable Bose gas has positive compressibility, indicated
by $\partial \mu /\partial n>0$.  In the normal state, a Bose gas
with $a_s<0$  collapses when $\partial \mu /\partial n=0$.  In the
Hartree-Fock approximation, this condition can be rewritten as
\begin{align}
-\frac{k_{\rm B}T_{\rm C}}{g}={1\over 2}\int
\frac{\mathrm{d}^3k}{(2\pi)^3}  \operatorname{csch}^2 \left(
\epsilon_k-\mu' \over 2 k_{\rm B}T_{\rm C} \right).
\end{align}
The collapse temperature $T_{\rm C}$ is always higher than the
molecular BEC transition temperature $T_p$ \cite{Stoof94, Mueller00, Yin02}. In the limit of weakly-attractive interaction,
$n^{1/3}a_s\rightarrow 0^-$, the collapse temperature $T_{\rm C}$
approaches to the ideal atomic BEC temperature $T_a$,
\begin{align}
T_{\rm C}=T_a\left[1+ {16\pi \over
3\zeta^{4/3}({3\over2})}n^{1/3}|a_s|\right].
\end{align}
In contrast, a normal Bose gas with $a_s>0$ is always mechanically
stable.

\begin{figure}
\includegraphics[width=7.5cm]{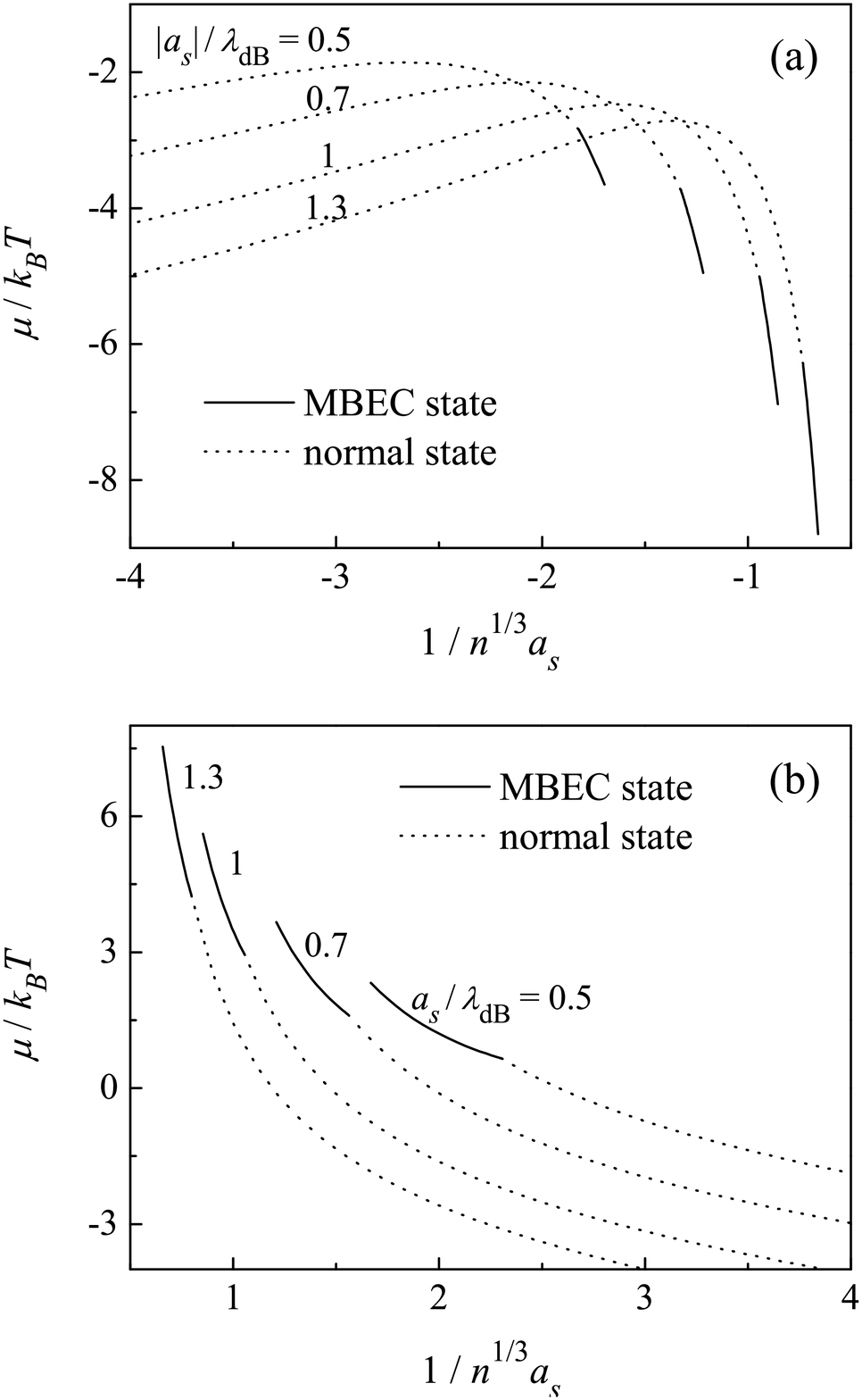}
\caption{Chemical potential $\mu$ versus $1/n^{1/3}a_s$.  The solid
line is the chemical potential in the molecular-BEC (MBEC) state and
the dotted line is the chemical potential in the normal state, (a)
with attractive interaction, $a_s<0$; (b) with repulsive
interaction, $a_s>0$. The lines with $|a_s|/\lambda_{\rm dB}=$0.5,
0.7, 1.0, and 1.3, are plotted, where $\lambda_{\rm
dB}=\sqrt{2\pi\hbar^2/mk_{\rm B}T}$ is the thermal de Broglie
wavelength.}\label{compress-pair}
\end{figure}

We examine the compressibility of the molecular-BEC state
numerically following Eq. (\ref{gapeq}) and (\ref{deneq}). In Fig.
\ref{compress-pair}, the chemical potential $\mu$ is plotted as a
function of density $n$ at given temperatures and scattering
lengths. For $a_s<0$, the chemical potential $\mu$ is monotonically
decreasing with the increase of $n$, $\partial \mu /\partial n<0$,
showing that the system is mechanically unstable as found in previous
studies \cite{Stoof94, Mueller00}. For $a_s>0$, the compressibility
is always positive, $\partial \mu/
\partial n>0$, and the molecular-BEC state is mechanically stable.
Therefore, in a uniform Bose gas, the molecular-BEC state can only
exist with positive scattering lengths, $a_s>0$.  In a trap, the
molecular-BEC state may be stabilized by the finite-size effect,
similar to trapped BEC with attractive interactions \cite{Hulet97}.

\subsection{Mean-field phase diagram}

The upper boundary of the molecular-BEC phase is determined by the
disappearance of order parameter $\Delta=0$, corresponding to the
second-order phase transition between the normal and molecular-BEC
phases.  At the transition temperature $T_p$, Eq. (\ref{gapeq}) is
the same as the Thouless criterion which marks the pairing
instability of the normal phase,
\begin{align} \label{Thouless}
-{1\over
g}=\int\frac{\mathrm{d}^3k}{(2\pi)^3}\left[\frac{1}{2(\epsilon_k-\mu')}
\coth\left({\epsilon_k-\mu' \over 2k_{\rm B}T_p}\right)-{1\over
2\epsilon_k} \right].
\end{align}
The density equation (\ref{deneq}) at $T_p$ becomes the Hartree-Fock
self-consistency condition of the normal Bose gas
\begin{align} \label{density-Tp}
n=\int \frac{\mathrm{d}^3k}{(2\pi)^3} {1\over
e^{(\epsilon_k-\mu')/k_{\rm B}T_p}-1}.
\end{align}

In the limit of the weakly-attractive interaction,
$n^{1/3}a_s\rightarrow0^-$,  the transition temperature $T_p$ is
close to the ideal atomic BEC temperature $T_a=(2\pi\hbar^2/mk_{\rm
B})[n/\zeta(3/2)]^{2/3}$,
\begin{align} \label{Tp_as0-}
T_p=T_a\left[1+{8\pi\over 3\zeta^{4/3}({3\over
2})}n^{1/3}|a_s|\right],
\end{align}
as found in a previous study \cite{Yin02}.

In the opposite limit, $n^{1/3}a_s\rightarrow 0^+$,  molecules are
tightly bound with the binding energy approximately given by
$E_{b0}$. From Eq. (\ref{Thouless}), the chemical potential is
approximately given by $\mu=-E_{b0}/2$.   The solution of density
equation (\ref{density-Tp}) yields
\begin{equation}
k_{\rm B}T_p=E_{b0}\ln^{-1}(E_{b0}/k_{\rm B}T_a)/3,
\end{equation}
which is essentially the same as the mean-field result about the BEC
limit of a Fermi gas with a BEC-BCS crossover. As pointed out in
earlier works \cite{Randeria93}, the divergent mean-field transition
temperature in this limit in fact corresponds to the temperature of
molecule dissociation, not molecular condensation.  The transition
temperature $T_p$ can be renormalized by pairing fluctuations
\cite{Stoof08}.

The lower boundary of the molecular-BEC phase is determined by
$\Delta=-\mu'$, where the excitation energy becomes gapless,
$E_k=\sqrt{\epsilon_k(\epsilon_k-2\mu')}$.  At zero temperature,
this boundary is located at $1/(n^{1/3}a_s)=4(3/\pi)^{1/3}\simeq
3.94$ \cite{Yin08}.  Disappearance of gap in excitation energies was
interpreted as appearance of atomic condensation \cite{Stoof94,
Stoof08}. However, as discussed later in section III, a coherent
mixture with atoms and molecules is mechanically unstable.

\begin{figure}
\includegraphics[width=8cm]{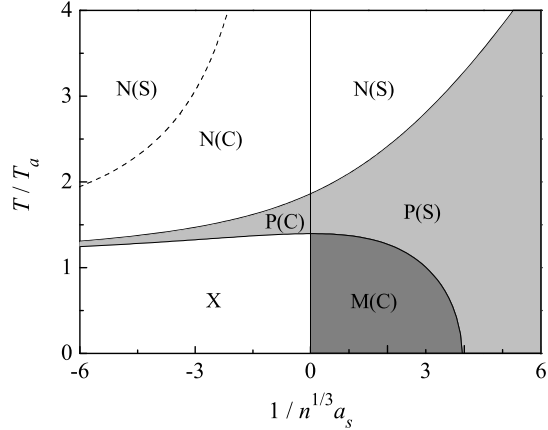}
\caption{Mean-field phase diagram of a pairing Bose gas with a wide Feshbach
resonance. P(S) denotes the stable molecular-BEC phase; P(C) denotes
the collapsing molecular-BEC phase; N(S) denotes the stable normal
phase; N(C) denotes the collapsing normal phase; M(C) denotes the
collapsing coherent mixture of atoms and molecules; X denotes no
solution according to the mean-field theory.  The Dashed line is the
mean-field collapse temperature of the normal phase with attractive
interactions.}\label{phase_diagram}
\end{figure}

The order parameter $\Delta$ in the molecular-BEC phase evolves
smoothly between two phase boundaries.  Based on the above results,
the mean-field phase diagram is plotted in Fig. \ref{phase_diagram}.
The molecular-BEC phase is labeled as P(S) for $a_s>0$ and P(C) for
$a_s<0$, where symbols (S) and (C) represent mechanical stability
and mechanical collapse.

Recently similar phase diagrams were obtained in Ref. \cite{Stoof08} within
the approach pioneered by Nozi\`{e}res and Schmitt-Rink (NSR) \cite{NSR85}.
In the NSR approach, pairing fluctuations are treated by the $t$-matrix
approximation, and the total particle density includes not only the mean-field
density, but also the density of thermal molecules.  Although there are some
discrepancies, most features of our mean-field
phase diagram Fig. \ref{phase_diagram} qualitatively agree with those
in Ref. \cite{Stoof08}.  The condensation temperature of the
MBEC phase is significantly reduced in the NSR approach. In the limit
of weakly-repulsive interaction,  it can recover the condensation
temperature of an ideal molecular Bose gas, $T_p=T_a/2^{5/3}$.
Both NSR and mean-field phase diagrams show that there is a transition
from the molecular BEC to the mixture state.  As shown in the next
section, the mixture state is mechanically unstable in the mean-field approximation.
In Ref. \cite{Stoof08}, the normal state was found to be stable in the simple NSR approach,
but unstable close to the resonance in the extended NSR approach
by taking into account interaction between molecules .
In the mean-field approach, the normal state is unstable near the
resonance only on the attractive-interaction side.

\section{Mechanical collapse of coherent atom-molecule mixture}

At the lower boundary of the molecular-BEC phase, the atomic
excitation energy vanishes, $E_0=0$, suggesting the appearance of
the coherent atom-molecule mixture. In the mixture, since both
molecules and atoms exhibit ODLRO, two order parameters can be
introduced \cite{Yin08}, $\psi_0\equiv\langle\psi\rangle$ describing
atomic condensate and $\Delta\equiv g\langle \delta\psi\delta\psi
\rangle$ describing molecular condensate, where
$\delta\psi\equiv\psi-\psi_0$. These two order parameters are in
principle independent, whereas in the atomic-BEC state $\Delta$ is a
function of $\psi_0$. In the mean-field approximation, the
Hamiltonian density for this mixture state is given by
\begin{align} \label{Hamilton-mixture}
\mathcal{H}_{m}=\mathcal{H}_m^{0} & -{\hbar^2\over
2m}\delta\psi^\dag\nabla^2\delta\psi
+(2gn-\mu)\delta\psi^\dag\delta\psi \nonumber \\ & +{1\over
2}\left[(\Delta+g\psi_0^2)\delta\psi^\dag\delta\psi^\dag+
h.c.\right],
\end{align}
where $\mathcal{H}_m^0=gn_0^2/2-\mu n_0-g\delta n^2-|\Delta|^2/2g$,
$n_0\equiv|\psi_0|^2$ is the atomic-condensate density, $\delta
n\equiv \langle \delta\psi^\dag\delta\psi\rangle$, and $n=n_0+\delta
n$.  The order parameter $\psi_0$ should minimize the mean-field
thermodynamic potential, which leads to the saddle point condition
\begin{align} \label{saddlepoint-mixture}
  \mu=g(n_0+2\delta n)+\Delta {\psi_0^*\over \psi_0} .
\end{align}
To satisfy the saddle point condition, the product $\Delta \psi_0^2$
must be real. For simplicity, we choose $\Delta>0$.

As in the molecular BEC case, the mean-field Hamiltonian of the
mixture can be diagonalized by Bogoliubov transformation, where the
field operator of atomic quasi-particles is given by $\alpha_{\bf
k}=u_k\psi_{\bf k}+v_k\psi_{-\bf k}^\dag$,
$u_k^2=v_k^2+1=[1+(\epsilon_k-\mu')/E_k]/2$, and the quasi-particle
energy is given by $E_k=
\sqrt{(\epsilon_k-\mu')^2-(g\psi_0^2+\Delta)^2}$.  The order
parameter $\Delta$ can be determined by the self-consistency
equation
\begin{align}
\Delta= -{g \over V} \sum_{\bf k}u_kv_k[1+2f_k], \nonumber
\end{align}
i.e.
\begin{align} \label{gapeq-mixture}
-{m\over 4\pi\hbar^2a_s}+{\psi_0^2\over \Delta+g\psi_0^2}= \int
\frac{\mathrm{d}^3k}{(2\pi)^3}\left[ \frac{1}{2E_k}\coth
\left({\beta E_k\over2}\right)-{1\over 2\epsilon_k} \right],
\end{align}
where similar to Eq. (\ref{gapeq}) the last term in the integrand is
the counter term. The equation for the total density is given by
\begin{align} \label{deneq-mixture}
n=n_0+\int \frac{\mathrm{d}^3k}{(2\pi)^3}\left[
\frac{\epsilon_k-\mu'}{2E_k}\coth \left({\beta
E_k\over2}\right)-{1\over 2}\right].
\end{align}

From the saddle point condition Eq. (\ref{saddlepoint-mixture}), the
ratio $\psi_0^*/\psi_0$ is either 1 or -1, and the quasi-particle
excitation energy can be rewritten as
$$E_k=\sqrt{(\epsilon_k+2gn_0)(\epsilon_k-2\Delta\psi_0^*/\psi_0)}.$$
If the excitation energy $E_k$ is real, two conditions, $gn_0>0$ and
$\psi_0^*/\psi_0=-1$, must be true, which cannot be satisfied when
the scattering length $a_s$ is negative, indicating that the mixture
cannot exist with $a_s<0$. When $\psi_0=0$, there is a transition
from the coherent mixture phase to the molecular-BEC phase. At zero
temperature, the transition point between the two phases locates at
$1/(n^{1/3}a_s)\simeq3.94$.

\begin{figure}
\includegraphics[width=8cm]{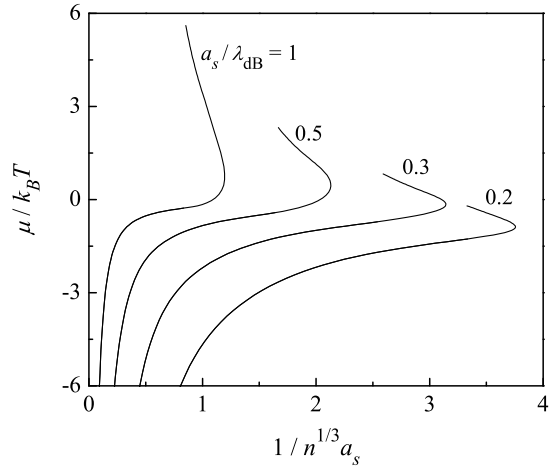}
\caption{Chemical potential $\mu$ of the coherent mixture versus
$1/n^{1/3}a_s$ with $a_s/\lambda_{\rm dB}=$0.2, 0.3, 0.5 and 1.0
respectively, where $\lambda_{\rm dB}=\sqrt{2\pi\hbar^2/mk_{\rm
B}T}$.  }\label{compress-mixture}
\end{figure}

For fixed total density, the chemical potential and order parameters
can be solved from Eq. (\ref{saddlepoint-mixture},
\ref{gapeq-mixture}, \ref{deneq-mixture}). The solution shows a
reentrant behavior near the transition to the molecular-BEC phase,
as shown in Fig. \ref{compress-mixture}.  The reentrance range in
temperature is broadened close to the resonance.  This reentrance
behavior is likely to be an unphysical result in the mean-field
approximation near the transition, similar to that in the Popov
approximation of a Bose gas near the atomic BEC temperature
\cite{Andersen04}.  Away from this reentrance, as shown in
Fig. \ref{compress-mixture}, the chemical potential $\mu$ is a
monotonically decreasing function of density $n$ for fixed $T$ and
$a_s$, $\partial\mu/\partial n<0$, meaning that compressibility is
negative.  Therefore the coherent mixture is subject to mechanical
collapse.

\section{Discussion and conclusion}

In the mean-field approximation, although the pairing gap and
Hartree-Fock energy are computed self-consistently, fluctuations
 are ignored.  In the NSR approach
\cite{Stoof08}, the pairing fluctuation is treated, which significantly reduced the condensation
temperature.  However, the Hartree-Fock
energy is ignored in the NSR approach, affecting the condition of
mechanical stability.  In future many-body theories, the self-energy should be
determined self-consistently with fluctuations properly treated.  Experimentally, three-body
recombination causes enormous particle-loss near the resonance.  In
a more accurate microscopic theory, both many-body and
few-body effects should be considered.
Nonetheless, the mean-field theory provides a basic picture about the strongly
interacting Bose gas, with most features of the phase diagram in qualitative agreement with the NSR
approach \cite{Stoof08}.

In conclusion, the molecular-BEC state of a homogeneous Bose gas
with a wide Feshbach resonance is studied at finite temperatures. In
the long-wavelength limit, the atomic excitation is gapped and
molecular excitation is gapless.  The molecular binding energy
is changed from its vacuum value due to many-body effect and
can even vanish when the scattering length $a_s$ is finite.
When $a_s<0$, the compressibility of the molecular BEC state is negative
and the system is subject to mechanical collapse.  The possibility of
coherent atom-molecule mixture is also explored, but the compressibility
of the mixture is negative and the mixture is subject to
mechanical collapse.  Based on these results, a mean-field phase diagram
at finite temperatures is obtained. This work is supported by NSFC under Grant
No. 10674007 and No. 10974004, and by Chinese MOST under grant
number 2006CB921402.

\end{document}